# Overview and main results of the DidaTab project

*François-Marie Blondel, Éric Bruillard, Françoise Tort*
*UMR STEF – INRP – ENS Cachan, UniverSud*
*61, avenue du Président Wilson*
*94235 Cachan cedex - France*
*francois-marie.blondel@inrp.fr*
*eric.bruillard@creteil.iufm.fr*
*francoise.tort@ecogest.ens-cachan.fr*

**ABSTRACT**

*The DidaTab project (Didactics of Spreadsheet, teaching and learning spreadsheets) is a three year project (2005-2007) funded by the French Ministry of Research and dedicated to the study of personal and classroom uses of spreadsheets in the French context, focussing on the processes of appropriation and uses by secondary school students. In this paper, we present an overview of the project, briefly report the studies performed in the framework of the DidaTab project, and give the main results we obtained. We then explore the new research tracks we intend to develop, more in connection with EuSpRIG. Our main result is that the use of spreadsheet during secondary education (grade 6 to 12) is rather sparse for school work (and even more seldom at home) and that student competencies are weak. Curricula have to be reviewed to include more training of dynamics tabular tools (including databases queries) in order to ensure sufficient mastery of computer tools that have became necessary in many educational activities.*

**1. INTRODUCTION**

As EuSpRIG points out for several years, uncontrolled and untested spreadsheet models pose significant business risks. To overcome such problem, many solutions can be foreseen, as regularly proposed in EuSpRIG conferences. One of them consists in improving education and training, for example developing spreadsheet error awareness. Underlying issues deal on the one hand with student spreadsheet competencies and on the other hand with effective uses of spreadsheets during secondary and post secondary or university education.

These issues were the main goals of the DidaTab project, funded by the French Ministry of research, during the last three years. The name of this project DidaTab means *Didactics of Spreadsheet* or *teaching and learning spreadsheets*, as Tab is for *tableur* which means spreadsheet in French. DidaTab is not directly a main concerns of EuSpRIG group, as it is not focussed upon professional uses of spreadsheets, but we think that some tools we have developed and some results we have obtained may relate to some issues dealt in EuSpRIG community and some of our research perspectives certainly are closer EuSpRIG interests. One hypothesis is that to get a broader view of users' competencies and their evolution may give interesting insights about what has to be done in the future to overcome badly designed spreadsheets.

In this text, just a few months after the official end of the project, we will try to give an overview of the project, its main goals and underlying ideas, give a list of what have been done and provide an executive summary of its main findings.

**2. DIDATAB PROJECT MAIN GOALS AND IDEAS**

Although ICT is included in the prescribed curricula in many European countries (Eurydice, 2004), very few data about effective practices in classrooms are available. Various indicators have been identified in national and international surveys (Lennon & al., 2003; Kvavik & Caruso, 2005; Demunter, 2006). But the ability to use professional software like spreadsheet is often based on self assessment, as in the survey of first-year university students by (Lim, 2005).

At least in France, IT competency of students is largely unknown and it is clearly an issue to determine if the students possess the necessary skills at the end of secondary education (grade 12). That was one of the objectives of the DidaTab project. More precisely, the initial idea was to:

– Provide an overview, as complete as possible, of spreadsheet uses in school and at home
– Underline differences in uses according to student profiles: kind of studies, gender, social milieu…
– Describe the various kinds of genealogy of use ("trajectories of use")
– Build a theoretical framework to explain the spreadsheet uses and competencies of students
– Identify the basic skills that students should be able to master
– Design exercises, questions and tests to assess these skills
– Evaluate students' skills using these tests

We had also the ambition to make some international comparisons and we organized research connections with French speaking colleagues from Belgium, Greece, Italy and Switzerland. This cooperation has been effective for curricula comparisons, for elaborating a common list of competences and for exchanging exercises and tests. A common work is in progress concerning interaction analysis using video transcripts.

As one can imagine, methodological issues were somehow tricky in this project with very limited duration (three years) and funding. How to describe trajectories of use without an extensive study when secondary education is seven years long and a lot of variables can potentially have an impact on spreadsheet uses and mastery (personal, historical or social characteristics as gender, social milieu, choice of stream of study, computer and mathematical skills…)? As explained in Bruillard & al. (2008), we have chosen a light approach, but well suited to our goals, relying on the help of Master students. We favoured polls or small surveys in a multiplicity of contexts; we crossed all these different studies in building a general framework, a model to link all the small results in specific contexts we got. We also used some tricks: as we had some hints at the beginning of the research (few uses of spreadsheets, lack of competencies), we selected sometimes most favourable situations according to spreadsheet uses (another reason is that it was not interesting for our project to investigate in places where nothing happens). So, as bias cannot be avoided, the results we obtained are surely better than the average situation.

We already have presented some partial results of our project (Blondel & Bruillard, 2006; Tort & Blondel, 2007; Haspekian & Bruillard, 2007; Bruillard & al., 2008), focussing each time upon specific points. We will focus here in the overall work done and main results obtained.

**3. WHAT HAS BEEN DONE?**

Our research work can be divided in five main components which correspond to different issues we had to deal with.

**3.1. Studies concerning prescribed and effective uses**

The first question to solve was to provide a detailed view of the uses of spreadsheet at home or in school. Eleven Master's dissertations have been produced, each one dealing with a specific subject matter (mathematics, physics, chemistry, economics, technology, engineering, management, trading, sport). Both official texts and educational resources (textbooks and websites) have been analyzed. 80 students have been interviewed on their uses of spreadsheet at home and at school. 30 teachers have been interviewed on their uses of spreadsheet in the classroom. Five complementary investigations have been carried out in several fields: interviews of professional users in companies (issue we want now to take more deeply into account as it was just for us a first spotting of spreadsheet uses in professional occupations), spreadsheet uses by blind students, spreadsheets and mathematics teaching, spreadsheets and biology and geology teaching ...

These different studies offer cartography, not a thorough picture but sufficient to give answers to the main questions we had.

**3.2. Analysis of spreadsheet competencies and identification of basic skills to be tested**

Another issue was to isolate elementary competencies in spreadsheet mastery, not related to specific applications as accounting. This task has been realised cooperatively in the framework of the DidaTab project with Monique Colinet (Namur, Blegium) and Etienne Vandeput (Liège, Belgium) and Vassilis Komis (Patras, Greece).

Every skill is described as a specific ability to use basic features to perform general but simple tasks. Based on these elements, the whole set of 87 basic skills is organized in 5 main categories and 23 sub-categories (see Tort and Blondel, 2006 for more details). Figure 1 gives the overall organisation.

| **Editing** | **Formulas** | **Graphs** | **Tables** | **Modelling** |
|---|---|---|---|---|
| Select Format Shape Copy Fill | Elements References Copy Errors | Data Types Present | Select Sort Extract | Identify Format Adapt Generate Print |

**Figure 1**: Spreadsheet competencies

For example, the first category (Cells and Sheets Editing and Formatting) covers: select and edit cells, format data types, copy and paste values, and auto fill. With 22 skills listed, this category is a basis for most of the others because it contains all basic manipulation of the objects displayed in a spreadsheet. Seven sub-categories have been designed:

- Select a cell or a range of cells
- Format a cell or a range of cells: number, date, currency…
- Change value appearance of a cell or a range of cells: font, bold, italic, borders …
- Apply conditional formatting to a cell or a range of cells
- Adapt cell appearance on a worksheet
- Copy cells with copy and paste
- Copy cells with auto fill

For another example, the last category (Modelling) covers: identify cell status, organize values, format and calculation, and structure for printing. It contains 13 skills with some high level abilities that we try to define independently of the application domain, like accounting, business or communication.

Every skill is described as a specific ability that can be linked to general but simple tasks (Fig.2).

| | |
|---|---|
| "Copy a cell or a range of cells into an adjacent or not adjacent zone" | (Editing) |
| "Translate a data processing described in natural language into a formula" | (Formulas) |
| "Use the effect of a copy on cells referencing in a formula" | (Formulas) |
| "Graph data with the suitable chart type" | (Charts) |
| "Choose a criterion for a one-key sort" | (Tables) |
| "Distinguish values obtained by calculation from others" | (Modelling) |

**Figure 2:** Examples of spreadsheets skills

### 3.3 Design of a collection of questions (paper and pencil) and tasks (computer-based).

The previous list of competence gives the basis of a collection of exercises: 150 paper questions and performance tasks have been built to serve as a basis for tests.

In a typical question, students are given a description and, if needed, a screen capture of a spreadsheet, and they are asked either what should be done to reach a given target, or what would happen if some actions are performed. In a typical performance task, students are given a spreadsheet that contains data from a very simple real life example. They must change or transform the appearance of the given spreadsheet (enlarge a column, bold some cells, sort columns) or make calculations from data (i.e. write adequate formulas as in Fig. 3), or create new objects, like charts. Figure 3 and figure 4 are examples of tasks.

**Figure 3**: Task "Write the sum calculation of a vector"

**Figure 4**: Change data display

In order to facilitate the design of tests, a documented database of skills, questions and tasks has been built. This database can be accessed through a web server. Tests designers can display skills, select tasks corresponding to a given criteria (author, tests, related skills), display the detailed information on a task, built and store information on new tests or new tasks.

| Title | Author | Difficulty | Medium |
|---|---|---|---|
| Automatic generation of dates | mc | 1 | computer |
| **Statement** | | **Solution** | |
| Enter the date of 'March 1st 2007' in cell A5 and then display the dates of all days in the month in the cells below | | Enter the date in the aaaa/mm/jj format, and copy it into the column below, using the auto fill handle | |
| **Skills** | | **Comment** | |
| Use the effects of copy and auto fill on the generation of series | | Setting a different cell format in cells below A5 allow viewing the effect of auto fill | |

**Figure 5**: A computer task in the DidaTab database

Our goal was to design tests, both paper and pencil and computer-based, for students enrolled in different streams. Each test must be consistent with the estimated level of skills and, allow differentiation among respondents' competencies.

A test consists in a series of performance tasks (or paper and pencil questions). Each task (or question) is related to spreadsheet skills that are relevant to the school level of the targeted population (junior high-school, senior high-school, tertiary or professional). A whole test covers all categories of skills (editing, formulas, charts, tables, modelling), and in a single test, there are several tasks (or questions) related to each category.

**3.4. Paper and computer tests**

We have built 3 paper tests intended respectively for low secondary school students, upper secondary schools students and students at university level. We have administrated these tests to 604 students. We have built 8 machine tests and have administrated them to 134 students**.** We have also recorded 36 student sessions during machine tests.

We gathered data from very different panels. With both paper tests and machine tests, we tested, in France: 121 low secondary school students (grade 7 and grade 9), 374 upper secondary schools students, and 61 university students; and also 165 students at university level in Belgium. These students were enrolled in different streams: general (humanities, science), technological (management, engineering), and vocational (business). As population we have tested has very various characteristics and according to the fact that it was not possible to control these variables, we will not give precise figures, but strong tendencies. We present numerical results in other papers (for example Tort and Blondel, 2007).

We have covered very different situations, giving very general results and some specific results. For example, we have chosen the class of a part time teacher trainer who has an interesting teaching practice adapted to non scientific secondary school students.

In-depth analysis of video recording is currently being done by our Greek partner.

### 3.5. Exploration of spreadsheet history and in-depth analysis of the spreadsheet literature

We have conducted a new analysis of the history of spreadsheet, based on very recent data (See http://hal.archives-ouvertes.fr/hal-00180912/fr/), opening our perspective concerning spreadsheets in education (See Baker & Sugden, 2003 for a classical review).

All theses research tasks contribute to build a comprehensive model of the role played by spreadsheet in French secondary schools. As previously explained in this paper, it is not possible to take into account all very specific cases one can encounter, but crossing all data we have gathered give some strong results and reasonable hypothesis about the use during secondary education and mastery of spreadsheet by students.

### 4. EXECUTIVE SUMMARY – KEY FINDINGS

The first key issue was the level of familiarity of students with spreadsheet. This level is weak in most of the cases. In average, students' mastering of spreadsheets is very low. Most of students are not able to efficiently use spreadsheets in their usual school activities.

### 4.1 Use of spreadsheets by students

Use of spreadsheets is sparse.

- Almost no use at home for personal needs, except for very few students in engineering (STI in French) or business / management (STG in French)
- Some uses at home for school work
- Most uses of spreadsheets occur at school.

These particular uses - that are not organised in time and occur at random (from a student point of view) - cannot give students the opportunity to build a sound knowledge of spreadsheets.

The episodic school activities involving spreadsheets are not sufficient to give students a clear view of what spreadsheets are and what can they be used for. These sparse uses are not sufficient for students to develop basic skills.

**4.2 Use of spreadsheets at school**

The use of spreadsheet at schools can be qualified as occasional or episodic. In lower secondary schools,

- Spreadsheets are taught during grade 7 in a subject called *Technology* (15 hours in average, mostly during practical activities in the computer lab);
- Some activities involving spreadsheets may be done in mathematic classes (grade 8 and grade 9), but the situation is certainly very different in one school from another.

In upper secondary schools, most of uses can be found:

- in maths and science courses: physics, chemistry, biology, geology, specially in labwork activities
- in courses in business/management, engineering, social sciences

*Prescribed national curriculum*

The prescribed national curriculum can be qualified foggy or fuzzy.

In the main streams of study - general and technological - the use of spreadsheets is often *recommended* and never compulsory. The recommendations put the emphasis on competencies rather than knowledge.

The official texts often assume that the basic spreadsheets skills are mastered by students so that they are able to use spreadsheets in school activities.

Some exceptions have to be noticed:

- The official text of the *Technology* course for the lower secondary level indicates precisely what should be taught about spreadsheets and listed the skills that should be mastered at the end of the course. It has to be noted that a new curriculum is under discussion where spreadsheets take a reduced place, even no place at all.

- Some courses in technical streams of study are directly related to desktop computer applications, where spreadsheet software has a part, for instance in the "social and medical sciences" (SMS in French).
- Spreadsheets have an explicit part in the maths course for students in the humanities stream (grade 12).
- Spreadsheets are also clearly mentioned in the prescribed curriculum of several tracks of the vocational stream (for office workers).

*Examination*

Spreadsheets are not used in assessment or final examination. So that it is not necessary for students to know something about spreadsheets to succeed in exams. The consequence is very easy to draw. For example in economical and social sciences, the prescribed curriculum gives some incitation to use spreadsheets with students, but it is not the case in most schools. For teachers, as final examination is a dissertation and computer use is not allowed, they consider that they have no time to spend using spreadsheets with students during school time.

Some exceptions:

The B2I certification which will be necessary to get the final exam at the end of lower secondary education (French *brevet des colleges,* grade 9) contains only one item (among 29) linked to spreadsheets (and only 23 items are required to get the certification).

Assessment of experimental skills that take place at the end of upper secondary school (French *baccalauréat,* grade 12) in physics, chemistry, biology, geology and also maths in a near future may rely on questions involving the use of a spreadsheet.

These exceptions may have an impact on teaching practices, especially in upper secondary schools.

*Masking spreadsheets*

Considering time constraints, even if they are convinced that spreadsheet software is useful for their students, teachers prefer not to spend time on teaching the basics of spreadsheets. If they finally include spreadsheets in their student activities, teachers manage to simplify the tasks in such a way that most of spreadsheets specificities and difficulties are wiped out.

As a consequence, these teaching practices have the effect of *masking* spreadsheets to students. In some cases, students are not able to recognize that they are using spreadsheet software when working on a particular task and even do not remember having used spreadsheets.

We observe a kind of vicious circle: as students have not enough competences, teachers tend to hide spreadsheet difficulties and, as students do not have to solve some difficulties, they do not acquire competences.

### 4.3 Competencies and skills

The analysis of the results of paper and computer tests show that high school students have low competencies and lack of confidence with spreadsheets.

Their best performances can be found in basic editing and formatting, skills that are common with other desktop applications like word processing. Specific features of spreadsheets like date/time formats or series generation remain unknown for a large majority of students.

Students seems to know the very basics of writing formulas, i.e. equals sign, arithmetic operators and relative cell reference, but they are able to use them only for very simple formulas. Most students do not use basic predefined functions like SUM() and are not able to manage the effect of copy/paste on relative cell references. Tests results are rather low on skills that require a deeper knowledge of the core of formula writing: syntax, predefined functions and absolute cells references.

Graphing skills are limited to vertical bar graphs (also named column charts in Excel), the default and also most popular type of graph in spreadsheet software. Few students seem capable to use other types of graphs, pie chart or scattered plot, for instance, when they are more suitable. And few students also are capable to associate several series of data in the same graph.

We have to notice some stream differences: students in the engineering and science streams show better performance in formula writing. Students in the business/management stream show better performance in editing and formatting.

### *Maths and spreadsheets*

The low confidence in spreadsheets can be linked to a lack of algebra skills, except for students in science (Haspekian, 2005; Haspekian & Bruillard, 2007). For example, mastery of copy and paste features requires some understanding of incrementation and even if the notion of variable appears to be very specific with spreadsheet, formula syntax requires some algebraic competencies. It seems that intrinsic difficulties of spreadsheet concepts are not sufficiently taken into account in mathematics education, even in the school streams where mathematics objectives and views are connected to every day life. The students who are mostly supposed to use the spreadsheet due to their school profile are unfortunately those who are precisely blocked by their difficulties in algebraic domain (for instance students in the "social and medical sciences" stream). And, at the opposite, students with better algebraic skills, for instance those in the science stream, are often left learning spreadsheets by themselves as if mathematical background is sufficient. We also have some examples of teachers in non scientific stream that are able to use spreadsheet as a new opportunity to enter algebra (but it is clearly an exception, far from being the standard rule).

### *Learning to work with computer applications*

Observation of students working with spreadsheets shows that they generally perform tasks using systematic exploration and trial-and-error techniques.

Very few students have a sufficient mastery of spreadsheet software to directly apply the right technique to answer a question or perform a task. Most of the time, they search through the interface which command or button will perform the required transformation. Or, when a formula is needed, they try several writings until the syntax is accepted and the results look acceptable.

Guidance given through the user interface, which may be automatically provided without interaction with the user, sometimes hides the underlying difficulty. For instance, this case may occur with the automatic correction of syntax when a formula is ill-formed, or with the sort buttons that perform ascending or descending sorting without selecting the whole range or also with the "Chart wizard" that guides the user to the default graph type. It seems that students learn how to use spreadsheets by combining a systematic exploration of menus and buttons with trial-and-error.

**4.4 Depreciation or denial of spreadsheets? A new vision is needed**

During our investigations, some unexpected issues (for us) have emerged, indicating some kind of denial of spreadsheet software.

First of all, computer science researchers show little interest in spreadsheet software as if this software was not worth while to be considered, as if no interesting problem can be associated with spreadsheet.

Secondly, in organizations, spreadsheets users are often unknown and their work is often unrecognized. We met several typical examples, some with very sophisticated uses (a significant number of spreadsheets users are programmers, doing end-user programming), and others with very basic ones.

Finally, teaching spreadsheets is considered as useless and not necessary because learning occurs "naturally" by practicing. Our research demonstrates that this last statement is false. And, in French education, students receive some minimal training (grade 7) and then are supposed to encounter ICT usage in every subject during their scholarship. Our study shows that it simply does not work.

So, to take into account seriously spreadsheet software in secondary education appears to be an important issue. In Bruillard & al. (2008), we have explained the implications of our findings for ICT education in France. Just to say a few words, with new online spreadsheets tools arising on the web, the need of a computer culture will probably become more crucial for regular users and, among them, teachers.

According to Jack Goody (1977), lists and tables are associated with writing and are not products of oral culture. But they played a key role in our written culture. Computers offer dynamic tables that can play similar roles and we have to be aware of this cultural shift. New automatic forms of online data aggregation are nowadays available at a large scale. Free online offers open the market to new opportunities and ICT instrumentation via the world network may have cognitive effects that could be as significant as those explained by Jack Goody (op. cit.) about written language. Interactive manipulation of tables can play an essential role in education, and it is urgent to make study on this emergent phenomena.

We know from a lot of studies that lack of competence leads to errors in spreadsheets and loss of time (for example Panko, 2000) and this problem has been identified for quite a long time (Brown & Gould, 1987), thanks to EuSpRIG community to try to give solutions to this problem. Our contribution is that solutions can also be found in elaborating new curricula for secondary education (grade 6 to 12 education) and in changing view of stakeholders concerning spreadsheet software.

**5. CONCLUSION AND FUTURE WORKS**

DidaTab project was mainly focused on secondary school students. We have some results concerning students at university level, but no study is available about what is supposed to be taught concerning spreadsheets or what is effectively taught or used at university or in post-*baccalauréat* schools. Furthermore, what are the uses of spreadsheets in French companies? As far as we know, no study has been done on that subject. Some interesting works are certainly to be launched in France concerning this question.

With colleagues of Paris V University, we have decided to launch some surveys and we try to find other contacts to launch similar surveys in other contexts, not in financial or accounting business, but in scientific fields. We will also focus on statistics education and links between spreadsheets and statistics or the use of spreadsheets to understand and practice statistics in human sciences fields.

We also continue to analyze the video transcripts we have recorded during some computer tests. These records provide very useful information about student behavior using spreadsheet and give interesting hints for explaining mistakes we have observed. Even if spreadsheet error analysis has produced many results, new users, coming with a different relationship with computer technology, may make new types of errors.

To conclude, misuses of spreadsheet software is certainly a very complex problem. Among possible answers, as information technology spreads out continuously and a lot of communication devices are used by the new generation, we think that trying to make young students more aware of computer processing chains, to give them strong basis to master information and communication technology can be important. The new millennium learners, according to an OECD expression (see Pedró, 2006), will develop new competencies. Will it be sufficient to solve spreadsheets problem? We think that the answer is no, if training is not dramatically developed.